\documentclass[aps,pra,twocolumn,showpacs,groupedaddress]{revtex4}
\usepackage{graphics,wasysym,amsbsy}

\begin{document}

\title{Practical scheme for a light-induced gauge field in an atomic Bose gas}

\author{Kenneth J. G{\"u}nter}
\email{kenneth.guenter@lkb.ens.fr}
\author{Marc Cheneau}
\author{Tarik Yefsah}
\author{Steffen P. Rath}
\author{Jean Dalibard}
\affiliation{Laboratoire Kastler Brossel and CNRS, Ecole Normale Sup\'erieure, 24 rue Lhomond, 75005 Paris, France}

\date{\today}

\begin{abstract}
We propose a scheme to generate an Abelian gauge field in an atomic gas using two
crossed laser beams. If the internal atomic state follows adiabatically the
eigenstates of the atom-laser interaction, Berry's phase gives rise to a vector
potential that can nucleate vortices in a Bose gas. The present scheme operates
even for a large detuning with respect to the atomic resonance, making it applicable
to alkali-metal atoms without significant heating due to spontaneous emission. We test the validity of the adiabatic approximation by integrating the set
of coupled Gross-Pitaevskii equations associated with the various internal atomic
states, and we show that the steady state of the interacting gas indeed exhibits a
vortex lattice, as expected from the adiabatic gauge field.
\end{abstract}

%\pacs{03.75.Nt}{Other Bose-Einstein condensation phenomena}
%\pacs{03.65.Vf}{Phases: geometric; dynamic or topological}
%\pacs{37.10.Vz}{Mechanical effects of light on atoms, molecules, and ions}
%\pacs{03.75.-b}{Matter waves}

\maketitle

A major motivation of current research with atomic quantum gases is the prospect of
simulating the physics of condensed matter systems (see, for example,
\cite{Bloch2008}). One objective is to gain new insight into strongly correlated
states of matter such as the fractional quantum Hall effect exhibited by electrons
in a magnetic field \cite{Stormer1999}. To study such systems with cold atoms, one
needs to mimic a gauge field with which the neutral atoms interact as if they were
charged. A well-known method consists in setting the gas into rotation and
transforming to the co-rotating reference frame, where the physics is governed by
the same Hamiltonian that describes charged particles in a magnetic field. At the mean-field level the corresponding gauge field leads to the nucleation of vortices, which has been observed experimentally (see~\cite{Fetter2008a} and references therein). However, technical difficulties have so far prevented the strongly correlated regime from being reached, 
and it is worthwhile to contemplate alternative approaches.

\begin{figure}[b]
 \scalebox{0.5}{\includegraphics{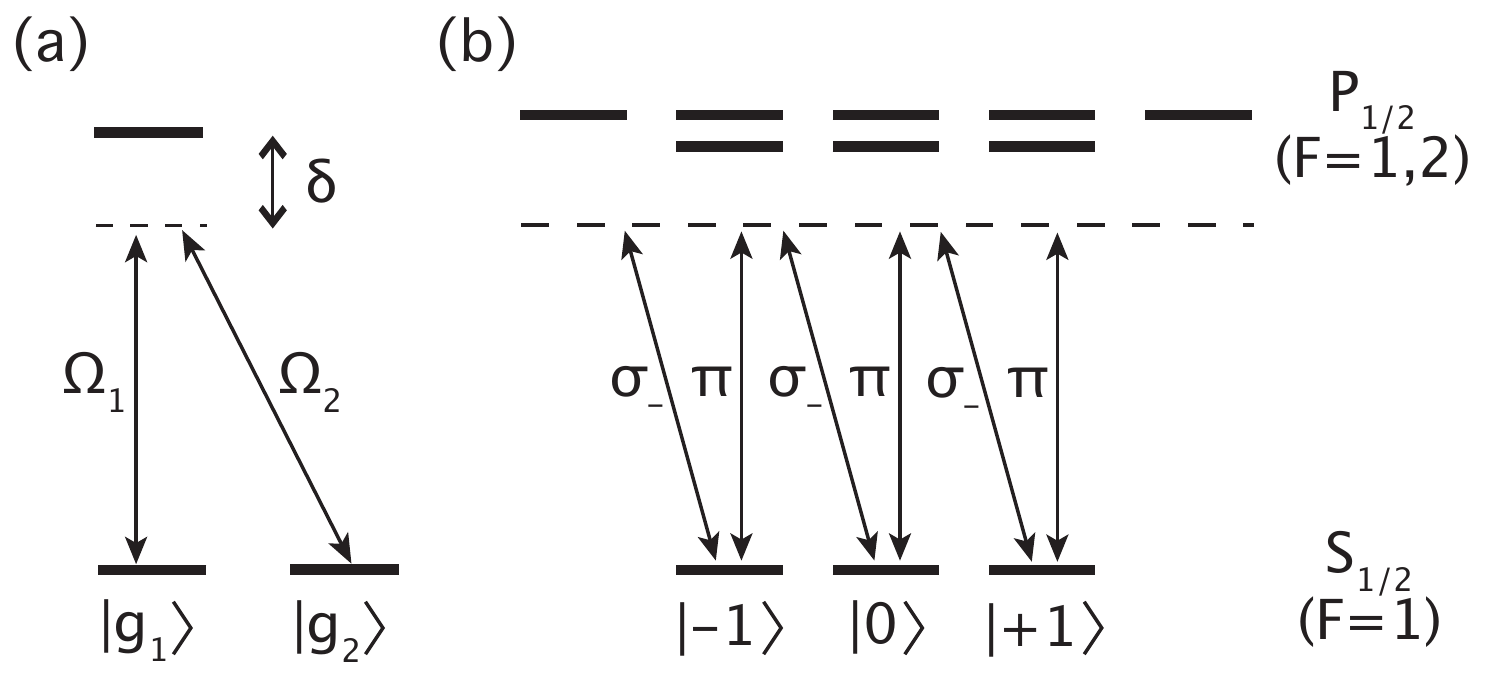}}
\caption{\label{figLevels} (a) $\Lambda$--type system used for a simple modelling of
the atom-laser coupling. (b) Atomic scheme relevant for a $3/2$ nuclear spin, as for
$^{87}$Rb or $^{23}$Na.}
\end{figure}

A promising possibility is the use of geometric potentials induced by laser fields. To comprehend the underlying mechanism, consider atoms with a ground state $g$, which is degenerate in the absence of any external perturbation (angular momentum $J_g>0$). Neglecting
in a first step the population of excited levels, the atomic energy eigenstates
$|\psi_\alpha\rangle$ ($\alpha=0,\ldots,2J_g$) in the presence of the laser fields are
linear combinations of the Zeeman substates $|g,J_g,m\rangle$ ($m=-J_g,\ldots,J_g$).
Both the eigenenergies $E_\alpha$ and the states $|\psi_\alpha\rangle$ depend on
position via the spatial variation of the light fields. If the centre-of-mass (COM)
motion is slow enough, an atom initially prepared in one of these internal states,
say $|\psi_0\rangle$, will remain in this state and its COM wave function will
acquire a geometric phase \cite{Dum1996}. This so-called Berry
phase \cite{Berry1984} corresponds to a gauge field $\{\boldsymbol{A},U\}$ appearing
in the effective Hamiltonian of the COM motion:
\begin{equation}
H=\frac{\left[\boldsymbol{p}-q\boldsymbol{A}
(\boldsymbol{r})\right]^2}{2M}+E_0(\boldsymbol{r})+U(\boldsymbol{r}),
\label{eqHamGauge}
\end{equation}
where $M$ is the atomic mass,
\begin{equation}
\boldsymbol{A}= i\hbar\langle \psi_0|\boldsymbol\nabla(|\psi_0\rangle), \quad U =
\frac{\hbar^2}{2M}\sum_{\alpha\neq 0}|\langle \psi_0|\boldsymbol\nabla(|\psi_\alpha
\rangle)|^2\ ,
\label{eq:potentials}
\end{equation}
and with the charge $q=1$. Several configurations to create gauge fields
with laser beams have been proposed and analysed
\cite{Jaksch2003,Juzeliunas2004,Mueller2004a,Sorensen2005,Zhang2005,Juzeliunas2005,Osterloh2005,Juzeliunas2006,Cheneau2008}, and
experimental evidence of these potentials has been provided in the context of cold
atomic gases \cite{Dutta1999,Lin2008}. In particular, the proposal of
\cite{Juzeliunas2006} uses two counterpropagating beams driving the two transitions
of an atom with a $\Lambda$ level scheme [Fig.~\ref{figLevels}(a)]. Such a
configuration emerges when the two beams are circularly polarised with positive and
negative helicity, respectively, and both the ground ($g$) and the excited ($e$) states
have unit angular momentum ($J_g=J_e=1$) \cite{Arimondo1996}. In this case
$|\psi_0\rangle$ can be chosen as the ``dark" superposition state of $|g,J_g,m=\pm
1\rangle$. Unfortunately, this scheme is not applicable to the commonly used alkali-metal
atoms even though they may have some hyperfine states with the proper angular
momenta. The hyperfine splitting of the relevant excited manifold being
typically smaller than 100 linewidths, non-resonant coupling to other
hyperfine states with $J_e\neq 1$ will destabilise the superposition state, and
heating due to spontaneous emission will eventually mask the effect searched for.

In this paper we first propose a simple planar scheme which overcomes
the above mentioned handicap. It incorporates two crossed laser beams whose
frequencies are far-detuned from the atomic resonance transition, which reduces the
rate of spontaneous emission processes to a negligible value. Our scheme produces an
effective gauge field that varies smoothly over the area of the atomic cloud and
allows for the observation of several vortices in a cold interacting spinor Bose
gas. In a second step, we determine the ground state of the gas by solving
numerically the coupled set of Gross-Pitaevskii equations (GPEs) associated with the
various internal (ground and excited) atomic states. We show that the ground state
of the system indeed contains several vortices with the predicted surface density.
In contrast to previous approaches \cite{Murray2007} where the gauge fields
(\ref{eq:potentials}) were explicitly included in the Hamiltonian, we do not {\it a
priori} assume the adiabatic limit to be valid. Hence, our study can be considered as
a validity check of this approximation for an interacting gas. In the last part of
this paper we investigate whether the vortices can be nucleated by slowly turning on
the geometric light potential.

The necessity of using off-resonant lasers strongly limits the choice of schemes that lead to a nontrivial gauge field. To be explicit, we first remark that a frequency detuning large compared to the hyperfine splitting of the excited manifold causes the nuclear spin to become irrelevant for the description of the excited state. Consider for example the $D_1$ transition between the states $S_{1/2}$ and
$P_{1/2}$ of an alkali-metal atom, which is sketched in Fig.~\ref{figLevels}(b) for the case
of a nuclear spin $3/2$, as for $^{87}$Rb or $^{23}$Na. Since only the
electronic angular momentum is relevant for evaluating the matrix elements of the
atom-laser coupling, the energy level structure can be regarded as a simple $J_g=1/2
\leftrightarrow J_e=1/2$ system. At the same time this restricts the polarisations
of the laser beams that can be applied to create the internal superposition state: A
spin flip from $m=-1/2$ to $m=+1/2$, for instance, corresponds to an angular
momentum transfer of $+\hbar$, so a two-photon transition requires light with
circular ($\sigma_-$ or $\sigma_+$) as well as with linear ($\pi$) polarisation.
Such a configuration cannot be achieved with counterpropagating waves as in
\cite{Juzeliunas2006} for which the change $\Delta m=0,\pm 2$
\footnote{Counterpropagating beams can induce a $\Delta m=\pm 1$ transition if a
comparatively large magnetic field is applied and the frequencies of the two beams
are properly shifted. Yet, the non-linear Zeeman effect may in that case lead to
spurious exothermic spin-changing collisions.}. Note that it is important that
the detuning is not large compared to the \emph{fine} structure splitting between the
$P_{1/2}$ and $P_{3/2}$ manifolds. Otherwise the electronic spin $S$ loses its
significance, and one is left with an effective $J_g=0\leftrightarrow J_e=1$
transition where no coherent level superposition is generated by the atom-light
interaction.

For the sake of simplicity, we first treat the case of a three-level
$\Lambda$ system. The transposition to a realistic internal level structure will be discussed later on. The
two ground states $|g_1\rangle$ and $|g_2\rangle$ are coupled to the
excited state by laser fields with spatially varying Rabi frequencies
$\Omega_1(\boldsymbol{r})$ and $\Omega_2(\boldsymbol{r})$, respectively
[Fig.~\ref{figLevels}(a)]. The Hamiltonian for an atom in the light field at any
point $\boldsymbol r$ reads
\begin{equation}
H=\frac{\boldsymbol{p}^2}{2M}+V(\boldsymbol{r}) +
H_\mathrm{AL}(\boldsymbol r). \label{eqHam}
\end{equation}
The atom is confined in a two-dimensional trapping potential
$V(\boldsymbol{r}) =(\omega_x^2x^2+\omega_y^2y^2)/2M$. The part
$H_\mathrm{AL}$ of the Hamiltonian acts on the internal degrees of freedom
only and describes the atom-laser coupling. Using the rotating wave approximation,
it can be written as
\begin{equation}
H_\mathrm{AL} =\hbar\left(\begin{array}{ccc} \label{eqHamLS}
0 & 0 & \Omega_1/2 \\
0 & 0 & \Omega_2/2 \\
\Omega^\ast_1/2 & \Omega^\ast_2/2 & -\delta
\end{array} \right),
\end{equation}
where $\delta=\omega_\mathrm{L}-\omega_\mathrm{A}$ is the detuning of the laser frequency $\omega_\mathrm{L}$ with respect to the atomic resonance frequency $\omega_\mathrm{A}$.
We suppose that the atom is prepared in the eigenstate
\begin{equation}
|\psi_\mathrm{0}(\boldsymbol r)\rangle =
\cos{(\theta/2)}|g_1\rangle+\mathrm{e}^{-i\phi}\sin{(\theta/2)} |g_2\rangle
\label{eqPsiNC}
\end{equation}
of $H_\mathrm{AL}$, where we have set $\cos{\theta}=(|\Omega_1|^2-|\Omega_2|^2)/\Omega^2$,
$\mathrm{e}^{i\phi}\sin{\theta}=-2\Omega_1^\ast\Omega_2/\Omega^2$, and
$\Omega=(|\Omega_1|^2+|\Omega_2|^2)^{1/2}$. Here $|\psi_\mathrm{0}\rangle$ is a
non-coupled (dark) state, separated from the next eigenstate by an energy given by
$\varepsilon=\hbar\Omega^2/4\delta$ in the low intensity limit $\Omega \ll |\delta|$.
The general expressions (\ref{eq:potentials}) for the gauge potentials yield in this case
\begin{eqnarray}
\boldsymbol{A}(\boldsymbol r) &=&
\frac{\hbar}{2}(1-\cos{\theta})\boldsymbol{\nabla}\phi, \label{eqVecpotGen}
\\
U(\boldsymbol r) &=&
\frac{\hbar^2}{8M}\left[(\boldsymbol{\nabla}\theta)^2+\sin^2\theta\,(\boldsymbol{\nabla}\phi)^2\right].
\label{eqScpotGen}
\end{eqnarray}

\begin{figure}[tb]
\scalebox{0.6}{\includegraphics{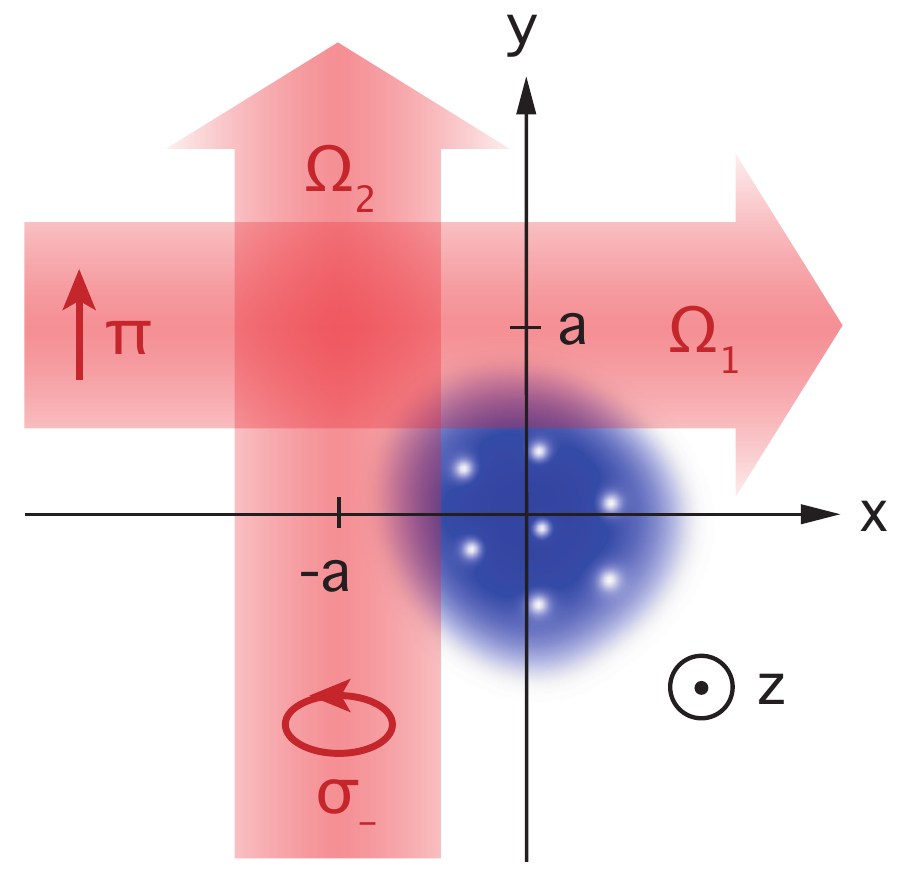}} \caption{\label{figScheme}Planar
scheme to create a gauge field for atoms with two crossed far-detuned laser beams
which are displaced by a distance $\pm a$ from the centre position of the atomic
cloud.}
\end{figure}

We now introduce a simple implementation scheme to generate a gauge field with two
far-detuned Gaussian laser beams. As the beams will in practice need to be
$\sigma_{-}$ and $\pi$ polarised (or equivalently $\sigma_+$ and $\pi$), we choose a
crossed-beam configuration.  Our beams are displaced by a distance $\pm a$ (of the order of the beam waists) from the $x$-- and the $y$--axis, respectively, as shown in
Fig.~\ref{figScheme}, in order to obtain a non-vanishing magnetic field at the
origin. The Rabi frequencies thus read $\Omega_1(\boldsymbol
r)=\Omega_0\exp[ikx-(y-a)^2/w^2]$ and $\Omega_2(\boldsymbol
r)=\Omega_0\exp[iky-(x+a)^2/w^2]$, where $w$ denotes the waist of the beams and $k$
the wave vector. Expanding the elements of the matrix~(\ref{eqHamLS}) up to second
order around the origin and plugging the results into Eq.~(\ref{eqVecpotGen}), we
obtain the vector potential
$\boldsymbol{A}(\boldsymbol{r})=\hbar
k(\boldsymbol{e}_y-\boldsymbol{e}_x)\left[1/2-a(x+y)/w^2\right]$, yielding the
effective magnetic field to lowest order:
\begin{equation}
\boldsymbol{B}(\boldsymbol{r})=\boldsymbol{\nabla}\times\boldsymbol{A}=\frac{2\hbar k
a}{w^2}\boldsymbol{e}_z. \label{eqBSch}
\end{equation}
An important quantity characterising the effect of the magnetic field is the
cyclotron frequency $\omega_c=q|\boldsymbol{B}|/M$, with $q=1$ in the present case.
For our configuration we get $\omega_c=2\hbar k a/M w^2$ close to the origin. For comparison,
in a gas rotating with angular frequency $\Omega_\mathrm{rot}$ one has
$\omega_c=2\Omega_\mathrm{rot}$. In the situation where several vortices are present
in the ground state, the surface density of vortices is related to $\omega_c$ by
\cite{Fetter2008a}
\begin{equation}
\rho_v=\frac{M\omega_c}{2\pi\hbar}=\frac{ka}{\pi w^2}. \label{eqVordens}
\end{equation}
The typical radius of the atom cloud can be chosen such that $R\sim a \sim w$ so
that the number of vortices inside the cloud is $\pi R^2 \rho_v \sim ka$. Owing to
the fast phase variation $\sim k \gg 1/a$ of the light field we may thus
expect a significant number of vortices to be present in a steady state cloud. This number is ultimately limited by the higher-order terms in the expansion of the gauge field. If the displacement $a$ is chosen too large compared to the waist $w$, the magnetic field will be inhomogeneous and the scalar potential anharmonic.

To obtain the scalar potential, we note that the first term in
expression~(\ref{eqScpotGen}) is small compared to the second one
$\sim(\boldsymbol{\nabla}\phi)^2=2k^2$ and can therefore be neglected. An expansion
around the origin up to second order then leads to
\begin{equation}
U(\boldsymbol{r})=
\frac{\hbar^2k^2}{4M}-\frac{1}{2}M\omega_c^2\left(\frac{x+y}{\sqrt{2}}\right)^2 \ ,
\label{eqScpotSch}
\end{equation}
which reduces the trapping frequency along the axis $x=y$. This anisotropy can be
compensated for by adapting the initial trapping potential $V$.

\begin{figure}[b]
\scalebox{0.495}{\includegraphics{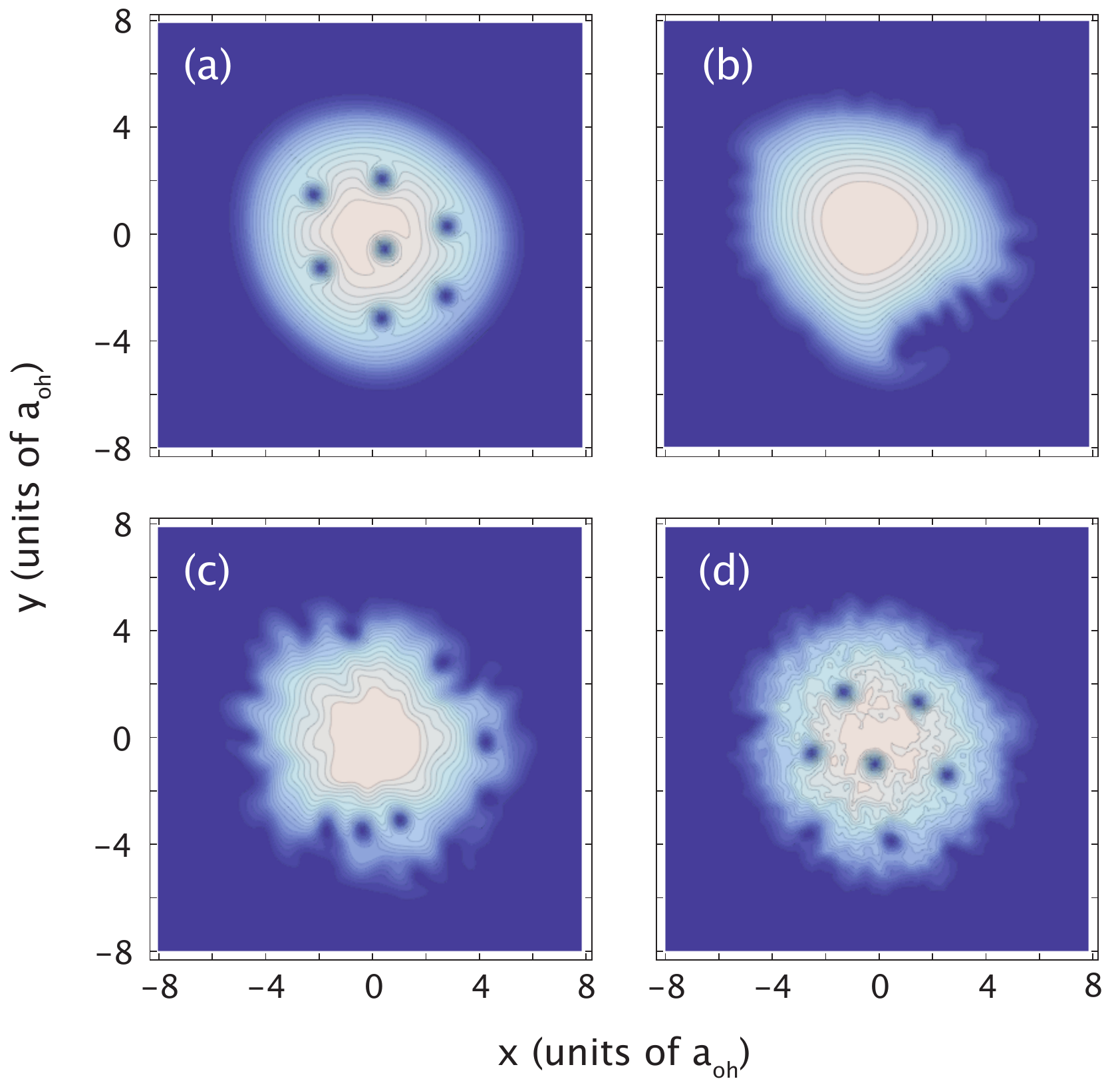}} \caption{\label{figDensad}
Density distributions of a two-dimensional (2D) gas of bosonic atoms with a $\Lambda$ level scheme
and rubidium-like parameters: resonance wavelength 795\,nm, $M=1.45\times
10^{-25}$\,kg, natural linewidth of the excited state $\Gamma/2\pi=6$\,MHz. The
harmonic trapping potential in the $xy$ plane, including the scalar potential $U$,
is isotropic with a trapping frequency $\omega/2\pi=40$\,Hz and a ground state
extension $a_{\rm ho}=(\hbar/M\omega)^{1/2}=1.7\,\mu$m. Atomic interactions are
chosen such that $G=Na_s(8\pi M\omega_z/\hbar)^{1/2}=800$. The laser
configuration is sketched in Fig.~\ref{figScheme}, with $a=w=25\,a_{\rm ho}$, $\Omega_0/\Gamma=280$, and $\delta/\Gamma=4\times 10^5$. (a)
Steady state distribution. (b)-(d) Distributions obtained after ramp
durations of the second laser beam of (b) $50\,\omega^{-1}$, (c) $200\,\omega^{-1}$
and (d) $1000\,\omega^{-1}$.}
\end{figure}

Up to now our scheme (like other previous proposals) relies on the assumption that the
atomic motion is slow enough so that the atoms follow adiabatically the state
$|\psi_0(\boldsymbol r)\rangle$. Although reasonable, this approximation could be
questioned in the vicinity of a vortex core, where the velocity field can become
arbitrarily large. To settle this point we have numerically determined the ground
state of a trapped gas of interacting bosonic atoms with a $\Lambda$--type
internal structure irradiated by lasers in the configuration of Fig.~\ref{figScheme}.
The gas is assumed to move only in the $xy$ plane, and the third motional degree of freedom
is assumed to be frozen to the ground state of a strongly confining potential along
the $z$ direction (frequency $\omega_z$). Spontaneous emission processes are
included by adding the complex term $-i\hbar\Gamma$ to the energy of the excited state.
For simplicity we take the same $s$--wave scattering length $a_s$ for all internal
states and hence disregard spin-spin interactions which could lead to spin exchange
collisions. This approximation is reasonable for the hyperfine ground states of
$^{87}$Rb but not necessarily for other atomic species. For the excited state it
should in general be unproblematic as the atoms spend their time primarily in the
ground state manifold. The interactions are treated at the mean-field level, and the
ground state of the system is obtained using an evolution in imaginary time of the three
coupled GPEs associated with the three internal atomic
states. A typical result is shown in Fig.~\ref{figDensad}(a). We find that the
atoms accumulate as expected in the non-coupled internal state $|\psi_0(\boldsymbol
r)\rangle$ while the population of states other than $|\psi_\mathrm{0}\rangle$ is
evanescent (fraction below $2\times 10^{-5}$ for the parameters of Fig.~\ref{figDensad}). Furthermore, the equilibrium COM distribution indeed exhibits a vortex lattice whose vortex density at the centre is in good agreement with the prediction of Eq.~(\ref{eqVordens}). The slight asymmetry of the cloud shape stems from non-harmonic contributions to the scalar potential (\ref{eqScpotGen}).

Vortex nucleation following the sudden application of a gauge field usually occurs
through a turbulent phase which lasts a fraction of a second, followed by a
relaxation to the ground state via a dissipative process
\cite{Madison2001,Tsubota2002,Lobo2004}. Another possible route could be
a smooth evolution from a no-vortex state to a state with vortices as the gauge
field is slowly turned on. Usually this process is forbidden, at least at the mean
field level, because of the different parities of the initial and final
states. However, the gauge field generated in the present scheme does not have any
definite parity property and such a smooth evolution should in principle be possible. To test whether it can be implemented in practice, we assume that one
laser beam is switched on first in order to lift the degeneracy between the two
ground state levels. Once it has reached its full intensity, we ramp up linearly the
intensity of the second laser beam in an adjustable time $T$. Starting with a
trapped gas in its ground state in the presence of one beam (a Bose-Einstein
condensate without vortices), we propagate the wave function in real time using the
GPEs. The resulting density distributions are shown in Fig.~\ref{figDensad}(bcd).
For short ramp times $T\lesssim 200\,\omega^{-1}$ many vortices are situated around
the border of the cloud but they have barely moved towards the centre [see
Figs.~\ref{figDensad}(b) and \ref{figDensad}(c)]. For slower ramps [Fig.~\ref{figDensad}(d)] the
final state looks closer to the expected ground state shown in
Fig.~\ref{figDensad}(a), although the convergence is not perfect. We conclude that
for these parameters the time scale for a smooth turn-on of the geometric potential
must be longer than $1000\,\omega^{-1}$ ($\approx4$\,s for $\omega/2\pi=40$\,Hz), which seems too long for a practical implementation. Probably vortex nucleation in these geometric potentials will be more easily achieved by following the same route as for a rotating trap, i.e., by suddenly applying the gauge field to a condensate at rest and by letting the cloud evolve to its new steady state through a transient turbulent phase [19-21].

%so it seems more likely that vortex nucleation in these geometric potentials should be
%obtained through the usual transient turbulent phase.

The analytical derivation presented above for a three-level system can be straightforwardly extended to the case of a realistic atomic transition. Although the results remain qualitatively unchanged, it is instructive to identify some minor
differences. Firstly, unlike for a $\Lambda$ system, the eigenstates of the
atom-laser coupling are not ``dark" in the sense that they all contain a non-zero
admixture of the excited level. However, the residual photon scattering rate can
be made very small. The parameters given in the caption of Fig.~\ref{figDensad}
lead to a scattering rate of 0.15\,photons/s per atom, which is negligible
on the time scale of a typical experiment. Secondly, we note that the level scheme
of Fig.~\ref{figLevels}(b) leads to some modifications of the geometric gauge field
with respect to the $\Lambda$ system. After optimisation of the relative intensities
of the $\sigma_-$- and $\pi$-polarised laser beams, one obtains that the effective
magnetic field at the origin is multiplied by the factor $8J_g/(3\sqrt{3})$ with
respect to the result~(\ref{eqBSch}). For the $J_g=1$ ground state of $^{87}$Rb, for
example, this leads to a 50\,\% increase of the vortex density at the trap centre.

In conclusion, we have proposed in this paper a scheme to create a
gauge field in a cold atomic gas with negligible heating due to photon
scattering. We have also validated the adiabatic approximation from which this gauge
field emerges by solving the set of coupled Gross-Pitaevskii equations associated
with the atomic internal levels. In particular, the ground state of a trapped 2D Bose
gas indeed contains several regularly arranged vortices, with the density predicted
from the gauge field in the adiabatic approximation. The feasibility of our simple scheme should render possible 
an experimental observation of a light-induced magnetic field. 

\begin{acknowledgments}
K.~J.~G. and S.~P.~R. acknowledge support from the EU (contract IEF/219636) and from the
German Academic Exchange Service (DAAD, grant D/06/41156), respectively. This work
is supported by R\'egion Ile de France (IFRAF), CNRS, the French Ministry of
Research, ANR (projects Gascor and BOFL) and the EU project SCALA. Laboratoire
Kastler Brossel is a \emph{Unit\'e mixte de recherche} of CNRS, Ecole Normale
Sup\'erieure and Universit\'e Pierre et Marie Curie.
\end{acknowledgments}


\begin{thebibliography}{20}
\expandafter\ifx\csname natexlab\endcsname\relax\def\natexlab#1{#1}\fi
\expandafter\ifx\csname bibnamefont\endcsname\relax
  \def\bibnamefont#1{#1}\fi
\expandafter\ifx\csname bibfnamefont\endcsname\relax
  \def\bibfnamefont#1{#1}\fi
\expandafter\ifx\csname citenamefont\endcsname\relax
  \def\citenamefont#1{#1}\fi
\expandafter\ifx\csname url\endcsname\relax
  \def\url#1{\texttt{#1}}\fi
\expandafter\ifx\csname urlprefix\endcsname\relax\def\urlprefix{URL }\fi
\providecommand{\bibinfo}[2]{#2}
\providecommand{\eprint}[2][]{\url{#2}}

\bibitem[{\citenamefont{Bloch et~al.}(2008)\citenamefont{Bloch, Dalibard, and
  Zwerger}}]{Bloch2008}
\bibinfo{author}{\bibfnamefont{I.}~\bibnamefont{Bloch}},
  \bibinfo{author}{\bibfnamefont{J.}~\bibnamefont{Dalibard}}, \bibnamefont{and}
  \bibinfo{author}{\bibfnamefont{W.}~\bibnamefont{Zwerger}},
  \bibinfo{journal}{Rev. Mod. Phys.} \textbf{\bibinfo{volume}{80}},
  \bibinfo{pages}{885} (\bibinfo{year}{2008}).

\bibitem[{\citenamefont{Stormer et~al.}(1999)\citenamefont{Stormer, Tsui, and
  Gossard}}]{Stormer1999}
\bibinfo{author}{\bibfnamefont{H.~L.} \bibnamefont{Stormer}},
  \bibinfo{author}{\bibfnamefont{D.~C.} \bibnamefont{Tsui}}, \bibnamefont{and}
  \bibinfo{author}{\bibfnamefont{A.~C.} \bibnamefont{Gossard}},
  \bibinfo{journal}{Rev. Mod. Phys.} \textbf{\bibinfo{volume}{71}},
  \bibinfo{pages}{S298} (\bibinfo{year}{1999}).

\bibitem[{\citenamefont{Fetter}(2008)}]{Fetter2008a}
\bibinfo{author}{\bibfnamefont{A.~L.} \bibnamefont{Fetter}},
  \bibinfo{journal}{arXiv:0801.2952}  (\bibinfo{year}{2008}).

\bibitem[{\citenamefont{Dum and Olshanii}(1996)}]{Dum1996}
\bibinfo{author}{\bibfnamefont{R.}~\bibnamefont{Dum}} \bibnamefont{and}
  \bibinfo{author}{\bibfnamefont{M.}~\bibnamefont{Olshanii}},
  \bibinfo{journal}{Phys. Rev. Lett.} \textbf{\bibinfo{volume}{76}},
  \bibinfo{pages}{1788} (\bibinfo{year}{1996}).

\bibitem[{\citenamefont{Berry}(1984)}]{Berry1984}
\bibinfo{author}{\bibfnamefont{M.~V.} \bibnamefont{Berry}},
  \bibinfo{journal}{Proc. R. Soc. A}
  \textbf{\bibinfo{volume}{392}}, \bibinfo{pages}{45} (\bibinfo{year}{1984}).

\bibitem[{\citenamefont{Jaksch and Zoller}(2003)}]{Jaksch2003}
\bibinfo{author}{\bibfnamefont{D.}~\bibnamefont{Jaksch}} \bibnamefont{and}
  \bibinfo{author}{\bibfnamefont{P.}~\bibnamefont{Zoller}},
  \bibinfo{journal}{New J. Phys.} \textbf{\bibinfo{volume}{5}},
  \bibinfo{pages}{56} (\bibinfo{year}{2003}).

\bibitem[{\citenamefont{Juzeli\=unas and \"Ohberg}(2004)}]{Juzeliunas2004}
\bibinfo{author}{\bibfnamefont{G.}~\bibnamefont{Juzeli\=unas}}
  \bibnamefont{and} \bibinfo{author}{\bibfnamefont{P.}~\bibnamefont{\"Ohberg}},
  \bibinfo{journal}{Phys. Rev. Lett.} \textbf{\bibinfo{volume}{93}},
  \bibinfo{pages}{033602} (\bibinfo{year}{2004}).
  
\bibitem[{\citenamefont{Mueller}(2004)}]{Mueller2004a}
\bibinfo{author}{\bibfnamefont{E.~J.} \bibnamefont{Mueller}},
  \bibinfo{journal}{Phys. Rev. A} \textbf{\bibinfo{volume}{70}},
  \bibinfo{pages}{041603(R)} (\bibinfo{year}{2004}).

\bibitem[{\citenamefont{S{\o}rensen et~al.}(2005)\citenamefont{S{\o}rensen, Demler,
  and Lukin}}]{Sorensen2005}
\bibinfo{author}{\bibfnamefont{A.~S.} \bibnamefont{S{\o}rensen}},
  \bibinfo{author}{\bibfnamefont{E.}~\bibnamefont{Demler}}, \bibnamefont{and}
  \bibinfo{author}{\bibfnamefont{M.~D.} \bibnamefont{Lukin}},
  \bibinfo{journal}{Phys. Rev. Lett.} \textbf{\bibinfo{volume}{94}},
  \bibinfo{pages}{086803} (\bibinfo{year}{2005}).

\bibitem[{\citenamefont{Zhang et~al.}(2005)\citenamefont{Zhang, Li, and
  Sun}}]{Zhang2005}
\bibinfo{author}{\bibfnamefont{P.}~\bibnamefont{Zhang}},
  \bibinfo{author}{\bibfnamefont{Y.}~\bibnamefont{Li}}, \bibnamefont{and}
  \bibinfo{author}{\bibfnamefont{C.~P.} \bibnamefont{Sun}},
  \bibinfo{journal}{Eur. Phys. J. D} \textbf{\bibinfo{volume}{36}},
  \bibinfo{pages}{229} (\bibinfo{year}{2005}).

\bibitem[{\citenamefont{Juzeli\=unas et~al.}(2005)
\citenamefont{Juzeli\=unas, \"Ohberg, Ruseckas, and Klein}}]{Juzeliunas2005}
\bibinfo{author}{\bibfnamefont{G.}~\bibnamefont{Juzeli\=unas et~al.}},
  %\bibinfo{author}{\bibfnamefont{P.}~\bibnamefont{\"Ohberg}},
  %\bibinfo{author}{\bibfnamefont{J.}~\bibnamefont{Ruseckas}}, \bibnamefont{and}
  %\bibinfo{author}{\bibfnamefont{A.}~\bibnamefont{Klein}},
  \bibinfo{journal}{Phys. Rev. A} \textbf{\bibinfo{volume}{71}},
  \bibinfo{pages}{053614} (\bibinfo{year}{2005}).
  
  \bibitem[{\citenamefont{Osterloh et~al.}(2005)\citenamefont{Osterloh, Baig,
  Santos, Zoller, and Lewenstein}}]{Osterloh2005}
\bibinfo{author}{\bibfnamefont{K.}~\bibnamefont{Osterloh et~al.}},
%  \bibinfo{author}{\bibfnamefont{M.}~\bibnamefont{Baig}},
%  \bibinfo{author}{\bibfnamefont{L.}~\bibnamefont{Santos}},
%  \bibinfo{author}{\bibfnamefont{P.}~\bibnamefont{Zoller}}, \bibnamefont{and}
%  \bibinfo{author}{\bibfnamefont{M.}~\bibnamefont{Lewenstein}},
  \bibinfo{journal}{Phys. Rev. Lett.} \textbf{\bibinfo{volume}{95}},
  \bibinfo{pages}{010403} (\bibinfo{year}{2005}).

\bibitem[{\citenamefont{Juzeli\=unas et~al.}(2006)\citenamefont{Juzeli\=unas,
  Ruseckas, \"Ohberg, and Fleischhauer}}]{Juzeliunas2006}
\bibinfo{author}{\bibfnamefont{G.}~\bibnamefont{Juzeli\=unas et~al.}},
  %\bibinfo{author}{\bibfnamefont{J.}~\bibnamefont{Ruseckas}},
  %\bibinfo{author}{\bibfnamefont{P.}~\bibnamefont{\"Ohberg}}, \bibnamefont{and}
  %\bibinfo{author}{\bibfnamefont{M.}~\bibnamefont{Fleischhauer}},
  \bibinfo{journal}{Phys. Rev. A} \textbf{\bibinfo{volume}{73}},
  \bibinfo{pages}{025602} (\bibinfo{year}{2006}).

\bibitem[{\citenamefont{Cheneau et~al.}(2008)
\citenamefont{Cheneau, Rath, Yefsah, G\"{u}nter, Juzeli\=unas, and Dalibard}}]{Cheneau2008}
\bibinfo{author}{\bibfnamefont{M.}~\bibnamefont{Cheneau et~al.}},
  %\bibinfo{author}{\bibfnamefont{S.~P.} \bibnamefont{Rath}},
  %\bibinfo{author}{\bibfnamefont{T.}~\bibnamefont{Yefsah}},
  %\bibinfo{author}{\bibfnamefont{K.~J.} \bibnamefont{G\"{u}nter}},
  %\bibinfo{author}{\bibfnamefont{G.}~\bibnamefont{Juzeli\=unas}},
  %\bibnamefont{and} \bibinfo{author}{\bibfnamefont{J.}~\bibnamefont{Dalibard}},
  \bibinfo{journal}{Europhys. Lett.} \textbf{\bibinfo{volume}{83}},
  \bibinfo{pages}{60001} (\bibinfo{year}{2008}).
  
\bibitem[{\citenamefont{Dutta et~al.}(1999)\citenamefont{Dutta, Teo, and
  Raithel}}]{Dutta1999}
\bibinfo{author}{\bibfnamefont{S.~K.} \bibnamefont{Dutta}},
  \bibinfo{author}{\bibfnamefont{B.~K.} \bibnamefont{Teo}}, \bibnamefont{and}
  \bibinfo{author}{\bibfnamefont{G.}~\bibnamefont{Raithel}},
  \bibinfo{journal}{Phys. Rev. Lett.} \textbf{\bibinfo{volume}{83}},
  \bibinfo{pages}{1934} (\bibinfo{year}{1999}).

\bibitem[{\citenamefont{Lin et~al.}(2008)\citenamefont{Lin, Compton, Perry,
  Phillips, Porto, and Spielman}}]{Lin2008}
\bibinfo{author}{\bibfnamefont{Y.-J.} \bibnamefont{Lin et~al.}},
 % \bibinfo{author}{\bibfnamefont{R.~L.} \bibnamefont{Compton}},
  %\bibinfo{author}{\bibfnamefont{A.~R.} \bibnamefont{Perry}},
  %\bibinfo{author}{\bibfnamefont{W.~D.} \bibnamefont{Phillips}},
  %\bibinfo{author}{\bibfnamefont{J.~V.} \bibnamefont{Porto}}, \bibnamefont{and}
  %\bibinfo{author}{\bibfnamefont{I.~B.} \bibnamefont{Spielman}},
  \eprint{arXiv:0809.2976} (\bibinfo{year}{2008}).

\bibitem[{\citenamefont{Arimondo}(1996)}]{Arimondo1996}
\bibinfo{author}{\bibfnamefont{E.}~\bibnamefont{Arimondo}},
  \bibinfo{journal}{Prog. Opt.} \textbf{\bibinfo{volume}{35}},
  \bibinfo{pages}{257–354} (\bibinfo{year}{1996}).

\bibitem[{\citenamefont{Murray et~al.}(2007)\citenamefont{Murray, \"Ohberg,
  Gomila, and Barnett}}]{Murray2007}
\bibinfo{author}{\bibfnamefont{D.~R.} \bibnamefont{Murray et~al.}},
  %\bibinfo{author}{\bibfnamefont{P.}~\bibnamefont{\"Ohberg}},
  %\bibinfo{author}{\bibfnamefont{D.}~\bibnamefont{Gomila}}, \bibnamefont{and}
  %\bibinfo{author}{\bibfnamefont{S.~M.} \bibnamefont{Barnett}},
  \bibinfo{journal}{arXiv:0709.0895}  (\bibinfo{year}{2007}).

\bibitem[{\citenamefont{Madison et~al.}(2001)\citenamefont{Madison, Chevy,
  Bretin, and Dalibard}}]{Madison2001}
\bibinfo{author}{\bibfnamefont{K.~W.} \bibnamefont{Madison et~al.}},
  %\bibinfo{author}{\bibfnamefont{F.}~\bibnamefont{Chevy}},
  %\bibinfo{author}{\bibfnamefont{V.}~\bibnamefont{Bretin}}, \bibnamefont{and}
  %\bibinfo{author}{\bibfnamefont{J.}~\bibnamefont{Dalibard}},
  \bibinfo{journal}{Phys. Rev. Lett.} \textbf{\bibinfo{volume}{86}},
  \bibinfo{pages}{4443} (\bibinfo{year}{2001}).

\bibitem[{\citenamefont{Tsubota et~al.}(2002)\citenamefont{Tsubota, Kasamatsu,
  and Ueda}}]{Tsubota2002}
\bibinfo{author}{\bibfnamefont{M.}~\bibnamefont{Tsubota}},
  \bibinfo{author}{\bibfnamefont{K.}~\bibnamefont{Kasamatsu}},
  \bibnamefont{and} \bibinfo{author}{\bibfnamefont{M.}~\bibnamefont{Ueda}},
  \bibinfo{journal}{Phys. Rev. A} \textbf{\bibinfo{volume}{65}},
  \bibinfo{pages}{023603} (\bibinfo{year}{2002}).

\bibitem[{\citenamefont{Lobo et~al.}(2004)\citenamefont{Lobo, Sinatra, and
  Castin}}]{Lobo2004}
\bibinfo{author}{\bibfnamefont{C.}~\bibnamefont{Lobo}},
  \bibinfo{author}{\bibfnamefont{A.}~\bibnamefont{Sinatra}}, \bibnamefont{and}
  \bibinfo{author}{\bibfnamefont{Y.}~\bibnamefont{Castin}},
  \bibinfo{journal}{Phys. Rev. Lett.} \textbf{\bibinfo{volume}{92}},
  \bibinfo{pages}{020403} (\bibinfo{year}{2004}).

\end{thebibliography}
\end{document}